\documentclass[useAMS,usenatbib,usegraphicx]{mn2e}
\usepackage{times}
\usepackage{url}

\title[Photo-z's from 2dFLenS ]{The 2-degree Field Lensing Survey: photometric redshifts from a large new training sample to $r<19.5$  }

\author[C. Wolf et al.]{Christian Wolf$^1$, Andrew S. Johnson$^{2,3}$, Maciej Bilicki$^{4,5}$, Chris Blake$^2$, A.~Amon$^6$, \newauthor
T.~Erben$^7$, K.~Glazebrook$^2$, C.~Heymans$^6$, H.~Hildebrandt$^7$, S.~Joudaki$^{2,3}$, D.~Klaes$^7$, \newauthor
K.~Kuijken$^4$, C.~Lidman$^8$, F.~Marin$^{2,3}$, D.~Parkinson$^9$, G.~Poole$^{10}$
\smallskip \smallskip \\
$^1$Research School of Astronomy and Astrophysics, Australian National University, Canberra, ACT 2611, Australia, email: christian.wolf@anu.edu.au \\
$^2$Centre for Astrophysics and Supercomputing, Swinburne University of Technology, PO Box 218, Hawthorn, VIC 3122, Australia \\
$^3$ARC Centre of Excellence for All-sky Astrophysics (CAASTRO) \\
$^4$Leiden Observatory, Leiden University, Niels Bohrweg 2, 2333 CA Leiden, The Netherlands \\
$^5$Janusz Gil Institute of Astronomy, University of Zielona Gora, ul. Szafrana 2, 65-516, Zielona Gora, Poland \\
$^6$Scottish Universities Physics Alliance, Institute for Astronomy, University of Edinburgh, Blackford Hill, Edinburgh EH9 3HJ, UK \\
$^7$Argelander-Institut f\"ur Astronomie, Auf dem H\"ugel 71, 53121 Bonn, Germany\\
$^8$Australian Astronomical Observatory, PO Box 915, North Ryde NSW 1670, Australia\\
$^9$School of Mathematics and Physics, University of Queensland, Brisbane, QLD 4072, Australia\\
$^{10}$School of Physics, University of Melbourne, Parkville, VIC 3010, Australia \\
}

\begin{document}
\date{draft \today}
\maketitle

\begin{abstract}
We present a new training set for estimating empirical photometric redshifts of galaxies, which was created as part of the 2dFLenS project. This training set is located in a $\sim $700 deg$^2$ area of the KiDS South field and is randomly selected and nearly complete at $r<19.5$. We investigate the photometric redshift performance obtained with $ugriz$ photometry from VST-ATLAS and W1/W2 from WISE, based on several empirical and template methods. The best redshift errors are obtained with kernel-density estimation, as are the lowest biases, which are consistent with zero within statistical noise. The 68th percentiles of the redshift scatter for magnitude-limited samples at $r<(15.5, 17.5, 19.5)$ are (0.014, 0.017, 0.028). In this magnitude range, there are no known ambiguities in the colour-redshift map, consistent with a small rate of redshift outliers. In the fainter regime, the KDE method produces $p(z)$ estimates per galaxy that represent unbiased and accurate redshift frequency expectations. The $p(z)$ sum over any subsample is consistent with the true redshift frequency plus Poisson noise. Further improvements in redshift precision at $r<20$ would mostly be expected from filter sets with narrower passbands to increase the sensitivity of colours to small changes in redshift.
\end{abstract}


\section{Introduction}

Redshift estimates for galaxies can be derived from imaging photometry and are known as photometric redshifts a.k.a. {\it photo-z's or phot-z's}. They were conceived over half a century ago to extend the reach of the largest telescopes in their attempt to constrain world models \citep{SW48,Baum57}. Today, they are particularly attractive for large-area surveys, where relatively modest observing time can deliver many more redshifts than spectroscopic campaigns. The motivation for photo-z's is still largely driven by cosmological tests \citep[e.g.][]{BB05,Mas15}, but extends beyond these to studies of galaxy evolution \citep[e.g.][]{Ben01,Wolf03a,Sco07,Spitler14} and the identification of rare objects.

Two domains of photo-z application can be differentiated: (1) Deep pencil-beam surveys, such as the original Hubble Deep Field \citep{HDF}, push the frontier of exploration into the unknown, and redshifts for distant faint objects are constrained by Bayesian exploration of the data using spectral energy distribution templates and galaxy evolution models \citep[e.g.][]{Lan96}. 

(2) At the other end of the scale, wide-area surveys grow to cover most of the sky and register huge numbers of galaxies despite relatively shallow flux limits, simply because area is easier to extend than depth; the analysis of their data is usually limited by systematics rather than number statistics. In this domain, photo-z's are ideally based on accurate empirical frequency maps of redshift occurrence, where such maps are usually derived from spectroscopic training samples such as that of the Main Galaxy sample of the Sloan Digital Sky Survey \citep[SDSS,][]{SDSS,FLS03,Csa03,Oya08a}. Photo-z catalogues of vast areas of sky have not only been constructed for the area covered by SDSS, but also for all-sky footprints, such as the 2MPZ and WISE x SuperCOSMOS catalogues by \citet{Bil14,Bil16}.

Given that photometric catalogues can easily achieve greater depth than complete spectroscopic catalogues, it is tempting to derive photo-z's as deep as $r>22$, although the available large and complete training sets such as the SDSS Main Spectroscopic Sample reach only $r\sim 17.5$. Since it is evident that fainter galaxies may reach higher redshifts, and empirical non-parametric maps cannot be meaningfully extrapolated, a deeper photo-z catalogue will only be useful when similarly deep training sets are added. This is now commonly done \citep[e.g.][]{Beck16}, although we note that most of these deep spectroscopic catalogues are highly incomplete \citep{New13}, and the objects with missing redshifts have been found to reside at very different redshifts when deeper spectra became available \citep{DeepPC}. Since the incompleteness propagates equally into the training sample as into the validation sample, it is not revealed by the purely internal performance measures of photo-z precision. When large parts of the true redshift distribution are missing from a training sample, they will be missing from the empirically trained photo-z catalogue as well as from the performance statistics. Pushing empirical photo-z's deeper in a reliable fashion requires not fancier statistical methods, but simply deeper complete random spectroscopic training samples.

In this paper, we explore photo-z's derived from a new training sample, which satisfies three important criteria for the first time: (1) going deeper than the SDSS Main Galaxy sample by two magnitudes, thus pushing to higher redshifts as well as fainter galaxies, (2) going wide enough to overcome cosmic variance by drawing the training sample from $\sim $700 deg$^2$ of sky, and (3) being very complete in representing the random galaxy population. Training sets can of course be derived from other surveys as well, however, e.g. the samples from the SDSS Stripe 82 \citep[e.g.][]{Nie09,Bundy15} and the WiggleZ survey \citep{Blake10,Dr10} use particular target selections and ignore certain types of galaxies. In contrast, redshifts from the Galaxy and Mass Assembly \citep[GAMA,][]{GAMA} survey are extremely complete and reliable to a similar depth as our new sample, and indeed manifest a complete census of the galaxy population within their target fields. The effective area of the GAMA sample used by \citet{Bil16} is, however, smaller than that of the sample presented here and may thus be more affected by cosmic variance. This is because GAMA fully samples its chosen sky area, while our training set subsamples a larger area. Thus, our new training sample should provide a useful resource for deriving galaxy photo-z's to nearly magnitude $r\approx 20$, and will be publicly available (see website at \url{http://2dflens.swin.edu.au}). It is an unbiased spectroscopic sample of general value, which can be used for photo-z training by any survey that covers the region of our sample.

The purpose of this paper is to explore how well broadband photo-z's perform at $r\la 20$ using complete random validation samples. The new training set was created with the SkyMapper Southern Survey \citep{K07} in mind, which will reach a depth similar to the SDSS imaging on 20,000 deg$^2$ area by 2019,
and release its first deep data soon \citep{W17}. The SkyMapper Southern Survey addresses a broad range of science goals: stellar science and Galactic archaeology studies benefit from the SkyMapper filter set ($uvgriz$), which allows estimating the stellar parameters $(T_{\rm eff}, \log g, M/H)$ straight from photometry. SkyMapper will be the main optical counterpart to the Evolutionary Map of the Universe \citep[EMU,][]{Norris11}, a large continuum survey of the Australian SKA Pathfinder \citep[ASKAP,][]{Jo07,Jo08} planned to run from 2017 to 2019. EMU will locate 70 million radio sources and will rely on photometric redshifts from SkyMapper, combined with VHS and WISE, for much of its work. Based on an average PSF with $2\farcs 5$ FWHM, we expect a point-source completeness limit of $>21$ mag in $g$ and $r$-band, so that SkyMapper will see counterparts to over 20 million EMU sources. SkyMapper will also be the main imaging resource to underpin the massive new spectroscopic Taipan Galaxy Survey at $i\la 18$ (see \url{http://www.taipan-survey.org}). Finally, the repeat visits of SkyMapper allow addressing variability in both the stellar and extragalactic regime. Most of the galaxies in SkyMapper will be at redshifts of $z<0.5$, and with SkyMapper being a legacy survey the photometric redshifts will be used for an unpredictable range of science applications.

In the absence of complete SkyMapper data, we based the photo-z exploration work on images of the slightly deeper VST-ATLAS survey \citep{Sh15}, where the filters are $ugriz$. Once SkyMapper data is available, we expect to see a slight improvement of redshift accuracy at the low-redshift end due to the extra violet filter of SkyMapper. As usual, we explore here not only empirical methods based on our new training set, but compare with a template method as well.

Our new training set is a complete random sample of galaxies from across a wide area of $\sim $700 deg$^2$ and obtained via spare-fibre spectroscopy within the 2-degree Field Lensing Survey \citep[2dFLenS,][]{Blake16}. The 2dFLenS survey is a large-scale galaxy redshift survey that has recently collected $\sim $70,000 redshifts within the footprint of the VST-ATLAS survey, using the AAOmega spectrograph at the Anglo-Australian Telescope. The principal science goal of 2dFLenS is to test gravitational physics through the joint observation of galaxy velocities, traced by redshift-space distortions, and weak gravitational lensing measured on data of the Kilo-Degree-Survey \citep[KiDS; see][]{deJong13,Kui15}. Its secondary purpose is to test methods of photometric-redshift calibration using both direct techniques for bright objects (this paper) and cross-correlation for a fainter sample \citep{J16}.

Our 2dF spectroscopy complements and extends existing data to increase depth and reliability of the photo-z's: the footprint of VST-ATLAS already contains several ten thousand published redshifts from the 2dF Galaxy Redshift Survey \citep[2dFGRS,][]{GRS}. While this data is complete and reliable to $r\approx 17.7$ obviating the need for bright-object spectroscopy, we limit our targets at the faint end to $r=19.5$, given the observing constraints of the 2dFLenS project and the methodological requirement to obtain a complete and reliable redshift sample. For all-sky photo-z purposes we desire a random sample that will represent the whole sky as best as possible and thus demand that its cosmic variance is as low as as possible. Hence, we construct our sample not as a complete census from a compact area, but instead by heavily subsampling the galaxy population across a wide area. The use of unallocated spare fibres in the wide-area spectroscopic survey 2dFLenS is thus a perfect solution for our needs. In the future, we will explore how to optimally include data from GAMA and other sources as well.

In the following, we present our data set in Sect.~2 and our methods in Sect.~3. In Sect.~4 we discuss the photo-z's we obtain by combining optical photometry in VST-ATLAS with mid-infrared photometry from WISE \citep{WISE} and by combining redshifts from 2dFGRS and 2dFLenS.

\section{Data}

The main objective of the 2dFLenS project is spectroscopic follow-up of the 1500 deg$^2$ VST-KiDS Survey \citep{deJong13}, which is optimised for weak-gravitational lensing studies. The target selection in 2dFLenS comprised several components with different requirements detailed in \citet{Blake16}. When 2dFLenS first started, KiDS did not yet cover all of its area, and therefore another imaging survey, VST-ATLAS \citep{Sh15}, was used in its place for 2dFLenS target selection. VST-ATLAS is less deep, but it covered the 2dFLenS area and was completely sufficient for source selection in 2dFLenS in terms of its depth and multi-band nature. For all details of survey coverage and processing of imaging data, we refer to \citet{Blake16}.

\subsection{Object selection from VST-ATLAS imaging data}

For the purpose of this paper, we only use relatively bright objects with $r<19.5$. We note that at the depth of VST-ATLAS, objects with $r=19.5$ have very low photometric errors in all bands (see Table~\ref{tab:ATLAS_charac}); in the $r$-band the median formal flux error is less than 1\%, and true uncertainties in the photometry are related almost exclusively to the usual challenges in galaxy photometry stemming from insufficiently constrained light profiles. 

 \citet{Blake16} describe the creation of our photometric catalogue for VST-ATLAS. In brief, we determine galaxy colours from isophotal magnitudes and use identical apertures in all bands. We first apply a shapelet-based PSF gaussianization and homogenisation \citep{K08,H12} to all images, then extract total object magnitudes in the detection r-band using flexible elliptical apertures, and run Source Extractor \citep{BA96} in Dual Image mode to obtain magnitudes for the other bands in consistent apertures. As a result, our galaxy photometry probes identical physical footprints on the galaxy image outside the atmosphere, despite bandpass variations of seeing. Finally, we apply illumination and standard dust corrections.  All magnitudes are calibrated to the AB system. We have also cross-matched the resulting catalogue to WISE (see Sect.~3.1.4 in \citet{Blake16}); there we find that most objects are detected in the W1 and W2 bands, while W3 and W4 have lower signal, and are thus ignored for this purpose. 
 
However, homogeneous photometric calibration turned out to be challenging in VST-ATLAS, because of the very low overlapping area between pointings (see Sect.~3.6 in \citet{Blake16}), and we noticed after creating our sample that the resulting mean photo-z biases were a function of VST field. We also found the mean object colours to vary in line with that, and hence need to apply additional zero-point offsets to remove the field-to-field variation in calibration. We thus modified the default photometry by adjusting the magnitude zero-points per VST field, using WISE as reference and adopting a very simple approach: (1) for every VST field we selected point sources with $r=[14,18]$, which is a nearly uncontaminated star sample with well-measured photometry; (2) we determined the median colours of this sample per field and compared with the overall median of the data set; finally, we (3) adjusted all bands so that the median colours per VST field are the same. In the latter step, we kept the WISE magnitudes as unchanged reference points, and only adjusted the VST-ATLAS photometry to match.

\begin{table}
  \caption{Average properties of VST-ATLAS imaging data in 2dFLenS area. $\sigma_{\rm mag}$ is the median magnitude error at the spectroscopic limit of $r=19.5$.}
\label{tab:ATLAS_charac}      
\centering          
\begin{tabular}{ccccc}
\hline\hline       
\multicolumn{1}{c}{Filter} & \multicolumn{1}{c}{exposure time [s]} &
\multicolumn{1}{c}{$\sigma_{\rm mag}$} & \multicolumn{1}{c}{seeing [$\arcsec$]} & $m_{\rm lim, 5\sigma}$ \\ 
\hline
$u$ & $2-4\times 60$ & $0.06$ & $1.11 \pm 0.20$ & 22.0\\ 
$g$ & $2\times 50$ & $0.02$ & $1.00 \pm 0.25$ & 23.0\\ 
$r$ & $2\times 45$  & $0.01$ & $0.89 \pm 0.19$ & 22.5\\ 
$i$ & $2\times 45$  & $0.01$ & $0.86 \pm 0.23$ & 21.8\\ 
$z$ & $2\times 45$ & $0.02$ & $0.87 \pm 0.22$ & 20.7\\
\hline                  
\end{tabular}
\end{table}

\subsection{Spectroscopy}

\subsubsection{The 2dFLenS direct photo-$z$ sample}

The spectroscopy for 2dFLenS was carried out using the AAOmega spectrograph at the Anglo-Australian Telescope between September 2014 and January 2016 (during the 14B, 15A and 15B semesters). In total, 2dFLenS contains about 70,000 high-quality redshifts at $z<0.9$, while this paper uses only the \textit{direct photo-$z$ sample}. This is a complete sub-sample of $\sim $30,000 galaxies in the magnitude range $17<r< 19.5$, that was added to the target pool of 2dFLenS at lower priority and has yielded spectra in the range $z\la 0.5$. 30,931 (31,864) targets were observed for the direct photo-z sample, for which 28,269 (29,123) good redshifts were obtained, where the figures in brackets include objects selected for the direct photo-z sample which were flagged for observation in other 2dFLenS target classes.

Starting from the photometric catalogues of VST-ATLAS, we selected all objects with $17<r<19.5$ as possible targets for this program. During the first two semesters (2014B and 2015A) of the 2dFLenS observations, we considered only clearly extended objects, requiring that the {\tt FLUX\_RADIUS} parameter in Source Extractor exceeded 0.9 multiplied by the seeing in all individual exposures and the co-added image. This selection produces a pure galaxy sample with extremely small stellar contamination, and ensures that little observing time is wasted on non-galaxies; however, it implies that the redshift sample does not represent the whole galaxy population randomly, but only the large portion identified as extended objects in our imaging data. Clearly extended galaxies are likely to have a different redshift distribution than compact galaxies, so this sample was intentionally incomplete.

During the last semester (2015B) of 2dFLenS observations, we included compact sources in the target list, in order to both, complement the galaxy sample and collect statistics on other types of unresolved sources with non-stellar colours. Using optical+WISE colours, it is possible to separate compact galaxies from stars, QSOs and host-dominated AGN \citep[see e.g.][]{J11,P16}, but a detailed investigation of this is beyond the scope of this paper. The number ratio between compact galaxies and extended galaxies with $r=[17,19.5]$ in our source catalogue is $1:12$.

Defining a redshift density distribution in any feature space works best when the redshift sample is both (1) representative, and (2) not sparse. While the representation criterion suggests a random sampling of galaxies, the sparsity criterion suggests to sample preferentially the low-density parts of the feature space. Their weight in the density estimation can be adjusted using appropriate selection weights, while their boosted number avoids the problem of discretisation noise. Given the steep number counts of galaxies, a pure random sample would be dominated by the faintest objects, while we would risk discretisation noise at the bright end. Hence, we decided to apply a magnitude-dependent weighting for the clearly extended galaxy sample to boost the brighter objects and moderately flatten the magnitude distribution in our observed sample: we used a factor $f(R)\propto 2^{(17-r)}$, which reduces the number count slope by 0.3.

\subsubsection{Redshifting}

Redshifts in 2dFLenS are measured and assigned quality flags using a combination of automatic software fitting and visual inspection.  First, all observed spectra are passed through the {\tt AUTOZ} code \citep{Bal14} developed for the Galaxy And Mass Assembly (GAMA) survey, which uses a cross-correlation method of redshift determination including both absorption-line and emission-line template spectra.  Each observed spectrum and {\tt AUTOZ} solution is then inspected by 2dFLenS team members using either the code {\tt RUNZ}, originally developed by Will Sutherland for the 2dFGRS, or the web-based {\tt MARZ} code \citep{MARZ}. The reviewers can then manually improve the redshift determination in cases where {\tt AUTOZ} is not successful and assign final quality flags.

Quality flags range from values of 1 to 5. Q=1 means that no features are visible, the redshift is purely determined by cross-correlation of the spectrum with templates and the correlation coefficient is very low. Only for Q=2 and better quality levels, a human assessor has actually labelled our confidence in the spectroscopic redshift as "possible" (Q=2), "probable" (Q=3), and "practically certain" (Q=4/5). We further group the sample into good-quality redshifts (Q=3/4/5) and bad-quality ones (Q=1/2).

\begin{figure*}
\centering
\includegraphics[clip,angle=270,width=0.95\hsize]{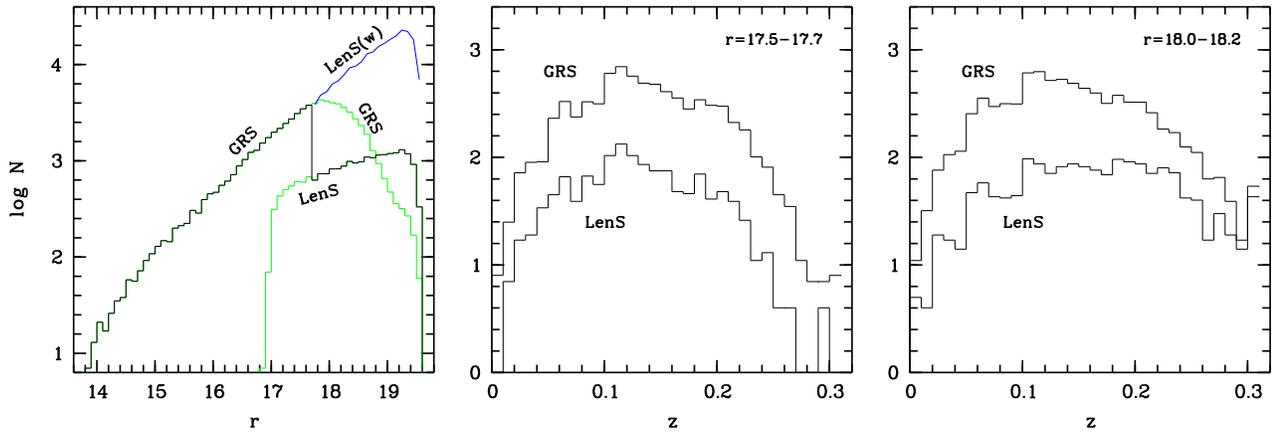}
\caption{Spectroscopic sample of empirical high-quality redshifts. {\it Left:} we combine 2dFGRS at $r<17.7$ with 2dFLenS at $r>17.7$ and give LenS galaxies a weight factor taking into account the larger effective area of GRS and our magnitude selection factor. The weighted number counts of the LenS sample connect seamlessly to GRS, so their combination forms a redshift sample with a realistic magnitude distribution. The edges of the LenS distribution are softened by a photometric recalibration after the sample was observed. {\it Centre:} Up to $r<17.7$ the redshift distributions of the shallower GRS and the deeper LenS agree, but fainter than that GRS appears to lack higher-z galaxies ({\it right}). Thus $r=17.7$ is a reliable completeness limit for GRS.
\label{mergedtrain}}
\end{figure*}

\subsubsection{Catalogue cleaning and completeness}\label{cleaning}

We construct our redshift sample for this paper from the 2dFLenS fields in the Southern and Northern area of the KiDS Survey. The Southern area spans the sky from $22^h$ to $3\fh5$ in R.A. and $-36\degr$ to $-26\degr$ in declination, and the Northern area extends from $10\fh4$ to $15\fh5$ in R.A. and $-5\degr$ to $-2\degr$ in declination. Thus, the combined sample covers a total effective area of $\sim $700 deg$^2$. We then remove (i) objects for which the photometry appears incomplete, flagged as unreliable or affected by artefacts, (ii) objects that are flagged by our image masks or have Source Extractor flags $>3$, which mostly eliminates objects with corrupted aperture data that are too close to the edges of images, and (iii) objects that are identified as Galactic stars from their spectra. We further eliminate objects for which the magnitude error reported by Source Extractor is 99~mag in any band; this indicates a faulty measurement as this value may appear for objects of any flux in the sample and does \textit{not} indicate a dropout. This cleaning process reduces our effective area without introducing a selection effect. 

Finally, the current version of the 2dFLenS redshift catalogue provides no reliable flagging of QSOs yet. For the purpose of this paper, we want to eliminate them from the sample, as they are rare and clear outliers from the main galaxy distribution in a magnitude-redshift diagram and cover their locus only sparsely. They are largely removed by applying a magnitude-dependent redshift cut to the 2dFLenS targets of $z<0.3+0.12(r-17)$. At the faint end of 2dFLenS this cut is at $z=0.6$. We take advantage of WISE photometry where available, but we do not require a WISE counterpart to use the object. In our model sample the fraction of galaxies without a WISE counterpart increases from 3\% at $r\sim 17$ to 13\% at $r>19$.

\subsubsection{Merging with 2dFGRS}\label{sample}

We combine redshifts from different samples to cover a broader range in magnitude, taking advantage of the fact that the 2dF Galaxy Redshift Survey (2dFGRS) and also the 6dF Galaxy Survey \citep[6dFGS,][]{6dF} already covered brighter magnitudes of $r\la17$ very well. Comparing the 6dFGS and 2dFGRS samples in the larger Southern field of the 2dFLenS area, we found 4,311 redshifts measured by both 2dFGRS and 6dF. Among these we find 29 disagreements ($\sim 0.7$\%), which appear to be mostly better measured by the deeper 2dFGRS. At very bright magnitudes of $14<r<15$ their completeness and quality is similar, while 6dFGS gets incomplete at $r>15.5$ and 2dFGRS extends well beyond $r=17$. We thus choose to build our master redshift sample simply from 2dFGRS and 2dFLenS, and select from the 2dFGRS just the $\sim $700 deg$^2$ region that extends from $22^h$ to $3\fh5$ in R.A. and $-36\degr$ to $-26\degr$ in declination and thus fully overlaps with the larger Southern area in 2dFLenS.

When merging two redshift samples of different quality, depth and effective area, we want to select and weight the objects such that they will appear to form a single, complete and unbiased sample taken from a consistent effective area. This is because the object mix in any training sample acts as a prior in empirical redshift estimation, and for optimum performance and lowest systematics we would like the prior to be unbiased. 

In Fig.~\ref{mergedtrain} we plot number counts of the two cleaned samples individually, which demonstrate that 2dFGRS is incomplete at $r\ga 18$, whereas 2dFLenS is by design incomplete at $r\la 17$ and $\ga 19.5$. The magnitude edges of the 2dFLenS distributions are soft, because we recalibrated the photometry again after the spectroscopic sample was selected. We also plot the redshift distribution $\log n(z)$ for both samples in narrow magnitude bins and find that 2dFGRS selectively lacks higher-redshift galaxies at $r\approx 18$. Hence, we cut the 2dFGRS sample conservatively to $r\le17.7$, where both number counts and redshift distributions compared to the deeper 2dFLenS suggest that it is entirely complete.

Since the sample definition of 2dFLenS was done prior to the final photometric calibration, we get just half a magnitude overlap where 2dFGRS and 2dFLenS are both complete. In this range, $\sim 80$\% of 2dFLenS galaxies were also observed by 2dFGRS, while 2dFLenS has observed $\sim 14$\% of the 2dFGRS targets owing to its much lower fill factor. From data of the first two 2dFLenS semesters, we found 1,274 high-quality redshifts measured in both surveys, with 14 disagreements (1.1\%) of $\Delta cz > 1000$~km/s. The vast remainder has a $\Delta cz$ RMS of 120~km/sec. Most of the 14 disagreements appear to be more reliable in 2dFGRS, and thus we decided to simply use 2dFGRS to $r\le 17.7$ and 2dFLenS at $r>17.7$.

Since the 2dFLenS sample is a spare-fibre sample with sparse coverage, its effective area is 17\% of that of 2dFGRS at $r=17.7$ (14\% over the Southern field alone, but 17\% overall once we add in the 2dFLenS targets from the Northern field). Furthermore, our selection boost for brighter, clearly extended, objects has flattened the number count slope by $-0.3$. Thus, we compensate the effective area of the 2dFLenS sample using magnitude-dependent weights of $\log w = -\log 0.17 + 0.3(r-17.7)$, where $w = 1$ for each 2dFGRS galaxy. The compact galaxies observed only in the third semester need $w = 20.41$ to represent them in the training sample with a weight corresponding to their abundance in the parent sample, from which all sample selection was done. 

The effect of these weights on the effective number counts of the 2dFLenS sample is shown in Fig.~\ref{mergedtrain} and appears to be a fair extrapolation of the 2dFGRS behaviour. The sample thus combined is a random sample of galaxies with $14\la r\la 19.5$, apart from incompleteness at the faint end due to edge-softening from recalibration and a mild decrease in the fraction of good-quality spectra that will be discussed later. It contains 50,919 good quality redshifts, of which 32,765 are from 2dFGRS and 18,154 are from 2dFLenS. We also build a sample of bad-quality redshifts (Q=1 or 2) with the same weight formula, which informs our magnitude-dependent redshift completeness: this includes a total of 2,191 bad redshifts, with 316 from 2dFGRS and 1,875 from 2dFLenS.

\section{Methods for photo-z determination}

From the earliest times of photo-z history, two different kinds of models have been used for photo-z's, those based on parametric templates and those based on empirical redshift data. Parametric models have redshift as one of their axes, while SED type, dust or star-formation histories are others. Empirical models are also known as training samples, even though not all empirical methods involve a true \textit{training} step. The empirical models simply sub-sample a discrete realisation of the true galaxy population within the survey with a size-dependent Poisson sampling noise and feature noise associated with the measurement process. Then they can be used as a model directly or feed a training step that creates an abstract model from the training sample. 

Either model defines a mapping $z(\vec{c})$ from the feature space, here photometry, to the label space, here redshift. For the empirical models, we can choose the feature space freely, while for template models they are restricted by the existing code package and its template information. Relevant criteria for choosing the feature axes are (i) maximising the redshift discrimination of features in the play-off between redshift dependence of the feature and typical noise for the feature values, as well as sometimes (ii) choosing features where the model is not too sparse and (iii) independence of features to minimise covariance. 

Features can be any observable not restricted to SEDs and can include e.g. size and shape parameters, in which case the estimates might not be called purely `photometric' redshifts. For the empirical models we chose to use a feature space spanned by the $r$-band magnitude, linking the optical and WISE using $r-W1$, and forming colour indices from neighbouring passbands otherwise; thus the full feature set is $\{r,r-W1,W1-W2,u-g,g-r,r-i,i-z\}$. The template method used here employs fluxes directly, and only the optical bands $ugriz$, because reliable templates that cover the mid-infrared wavelengths are not yet available.

All methods used in this paper are probabilistic, as opposed to function-fitting methods, so they provide redshift distributions $p(z|c)$ given a colour measurement. An important point worth clarifying in the context of photometric redshifts is the usage of the term probability. In his textbook about statistical inference \citet{MacKay} reminds us that \textit{probability} can describe two different meanings: it can describe ``frequencies of outcomes in random experiments'', and it can describe ``degrees of belief in propositions that do not involve random variables'', and further notes that ``a likelihood function is not a probability distribution''. This is especially relevant when photo-z catalogues are used in a cosmological analysis, where probability distributions are often taken to be frequency distributions. E.g., weak-lensing and clustering studies have long benefitted from considering the full probability redshift distributions of individual objects \citep[e.g][]{Edd06,Kit07,Man08,Kilb15,Asorey16}. Hence, we will also investigate towards the end of the paper to what extent the $p(z)$ distributions we obtain represent actual frequency distributions $n(z)$.

We also note a fundamental floor to the precision with which the redshift of a galaxy can be estimated from its SED, irrespective of method. The intrinsic variety in galaxy colours at a fixed redshift implies that for an observed galaxy colour there is intrinsic variety or ambiguity in redshift. While this effect is obvious when the observable is a single colour index, it still dominates the redshift errors of bright well-measured galaxies in current multi-band data sets. A few broad passbands don't break all degeneracies in the space of SEDs spanned by the entire observable galaxy population, and even with a very rich training set these intrinsic limits will not be overcome, but are set by the observed features. Indeed, the primary aim of creating the \textit{direct photo-z sample} in the 2dFLenS Survey was to create a model sample at $r\la 20$, which would allow the derivation of empirical photo-z's with a precision that approaches the theoretical limit. In contrast, e.g. \citet{PHAT} compare a large number of methods on a deep photometric dataset from the Great Observatories Origins Deep Survey \citep[GOODS,][]{Giav04} with sparser and incomplete spectroscopy, which is the extreme opposite of the data domain investigated here.

At bright magnitudes calibration uncertainties may be large relative to photometric errors, and will thus drive the error budget. For template methods this concerns the calibration of the model templates relative to the observations. The issue is equally important for empirical methods, where the sky area containing the model sample needs to be on the same calibration as the query sample. Otherwise, colour offsets between query and training data will lead to systematic redshift biases.

Our photometric data is much deeper than the spectroscopic data: errors are typically $<2$\% at the faint end of the spectroscopic sample, with the exception of the u-band, where the median error at the faint limit is $\sim $6\%. This means that calibration errors dominate the results in our study, and we will find the floor of the possible photometric redshift errors. Even deeper u-band observations are expected to improve the results only marginally.

\subsection{Empirical method KDE}

Kernel-density estimation \citep[KDE, see e.g.][]{Wang07} was one particular method, where the Bayesian and empirical approach could be unified by using the empirical sample objects as discrete instances of a model: technically KDE with an empirical sample is identical to Bayesian model fitting, if the kernel function is chosen to be the error ellipsoid of the query object and the empirical features of the model object are free of random errors, or the kernel function is the square difference of query and model errors \citep[see][]{W09}.

In practice, the KDE method runs over one query object at a time and determines the redshift probability function $p(z)$ at the location of the query object from a representation of all other model objects obtained by convolving the discrete point cloud with a kernel. As a kernel function we use a Gaussian, whose width is the squared sum of the photometric error of the query object and a minimum kernel width. Mathematically, this is identical to a template-fitting code that determines a likelihood from a $\chi^2$-fit that square-adds a photometric error and a minimum error to take calibration uncertainties into account. For each query object with the features $c_{\rm query},j$ we find a probability $p_i$ that it resembles a model object $m_i$ with the features $c_{i,j}$ and located at redshift $z_i$, which, assuming Gaussian errors, is derived by the standard equation 

\begin{equation}
	\chi^2_i = \sum_j \frac{(c_{{\rm query},j} - c_{i,j})^2}{\sigma_{{\rm query},j}^2 + \sigma_{0,j}^2}   ~,\\
	p_i \propto e^{-\chi^2_i/2}  ~.
\end{equation}

We choose the minimum kernel width to be $\sigma_{0,j}=0\fm 05$ for colour indices, and $0\fm 2$ for the $r$-band magnitude: this kernel smoothing is not meant to signify a calibration uncertainty, but to cover the sparsity of the model; and in analogy to a template-fitting method, smoothing over the $r$-band magnitude limits the resolution of a magnitude prior, while smoothing over a colour index limits the impact of the SED itself. Template-fitting methods commonly assume $\sim0\fm 05$ uncertainties on zeropoints, but they require no smoothing and thus no error floor for a magnitude prior, as their nature is not discrete but continuous.

We exclude the query object itself from the model, because otherwise it would appear as an identical pair in the above equation with $\chi^2_i=0$ and mislead the results. This form of query-object exclusion allows us to use the full empirical sample both as a query sample and as a model sample. Neural network training  (discussed in Sec. 3.2), in contrast, requires splitting the sample into separate non-overlapping training and validation samples, perhaps using a 70:30 split, in order to prevent overfitting. In the KDE method each object available in the model sample gives an independent estimate of redshift precision and the whole sample can be used as a model. However, increasing the model size from 70\% of a training sample to 100\% in a KDE model sample improves the estimation performance only slightly, but the over $3\times$ bigger validation sample reduces the noise in the validation result. The latter aspect does not make the estimation better, but increases our confidence in measuring the performance.

Naturally, this approach produces probabilities for discrete redshift values. We then resample these into redshift bins to cover the continuous redshift range, by sorting all model objects and their associated probabilities $p_i$ into bins of width $\Delta z = 0.003 \times (1+z)$. We set an upper redshift limit at $z=0.5$, since in our magnitude range, objects at higher redshift are very rare and populate the feature space only sparsely. We eliminate higher-redshift objects only from the model and thus preclude an assignment of a higher redshift estimate. However, we keep the few higher-redshift objects in the query sample, effectively forcing them to appear as outliers, since they would typically form a small but real part in any blind photometric sample subjected to photo-z estimation. 

The redshift probability distribution $p(z)$ is then normalised, where we take two possible classes of objects into account: (1) The first class is the empirical sample of high-quality redshifts as described in Sect.~\ref{sample}, for which we get a meaningful $p(z)$ distribution. (2) The second class comprises the remainder of the complete target sample, where no reliable redshift was obtained from the spectra, and these are attributed an "unrecoverable $p(z)$". Each object in the query sample is compared against both model classes, and their two Bayes factors are used to normalise the probability integrals. For each object we obtain the resulting "X\% probability" that the object is drawn from the redshift distribution $p(z)$, while with ($100-$X)\% probability it is drawn from an unknown distribution. The use of these explicit models allows the probability of an unknown redshift to be measured on a per-object basis, sensitive to the local completeness in its own region of feature space. Thus we can flag more easily which specific objects in the query sample have uncertain estimates. 

The code is currently not sufficiently documented to be published, but it has been used with pre-calculated template grids by \citet{W99} in CADIS and \citet{W04} in COMBO-17; it has also been used with empirical models for obtaining the more challenging photo-z estimates of QSOs from SDSS photometry \citep{W09}, where redshift ambiguities are more common than with galaxies at low and moderate redshift. There the PDFs were shown to be particularly successful in predicting true redshift frequencies and thus relative probabilities for alternative ambiguous solutions. It was further tested with a smaller and spectroscopically incomplete model sample at very faint magnitudes in the comparison by \citet{PHAT}, where it naturally could not play out its advantages. The code uses an implementation of the above equation in C that reads FITS tables with data and models, and uses wrapper scripts to handle the metadata for features and models.

\subsection{Empirical Method: Neural Networks}

Neural Networks (NNs) are a collection of {\it neurons} arranged into layers. In the simplest case there exists an {\it input layer}, a {\it hidden layer} and an {\it output layer}. Each neuron is connected to all the other neurons in the previous layer, and these connections are assigned a specific weight. As the NN `learns' from the training data these weights are adjusted. The output of a neuron is a scalar quantity, therefore, each neuron is a function which maps a vector to a scalar quantity. As the input vector reaches a threshold value the neuron {\it activates}, that is, the output value changes from zero to a non-zero value -- the biological parallel is the firing of a neuron in a brain. This process of activation is controlled by an activation function, in our case this is a $\tanh$ function. 

In order to obtain redshift PDFs from the network instead of single-value estimates, we design it as a probabilistic neural network \citep[PNN,][]{PNN}, and thus map the estimation problem to a classification problem, where the classes are fine redshift bins; a photo-z application of this type was published by \citet{Bon15}. We used the public code {\tt Skynet} \citep{2014MNRAS.441.1741G}, which was also tested in the Dark Energy Survey comparison by \citet{San14}. It returns a weight for each object and class and turns these weights into a probability distribution using a {\it softmax} transform. For the NN architecture we used three hidden layers with 30, 40, and 50 neurons per layer. The input is 10 variables (magnitudes and colours) and the output is the redshift PDF divided into 50 separate redshift bins. We chose 50 bins from $z=0.002$ to $z = 0.40$ and chose the truncation at $z=0.4$, because the code could not handle the limited number of higher redshifts available. Also, with a constant bin width the number count in each class would vary significantly and cause the accuracy to vary with redshift. To avoid this, we adjust the width of each bin such that the number count distribution is as uniform as possible, but we cap the width at a maximum of $\Delta z=0.04$. As a result we have $N \sim 1100$ in each bin. Note that, when weights are introduced into the training process, the bin sizes are adjusted to make the weighted number count uniform.

In order to train a network, we need to break the spectroscopic dataset into a {\it training} and {\it testing} sample. For our case, the testing sample will also act as the {\it validation} sample. The training sample is used to infer the mapping from feature space to label space, while the independent testing sample is used to evaluate the performance of the redshift estimates. However, a problem with NNs is their sensitivity to noise, i.e. statistical fluctuations in the data. This occurs when network architectures are overly complicated with too many neurons. We note this problem is avoided in ${\tt Skynet}$ because of the convergence criteria implemented. While the network is training, the algorithm computes the sum of the squared redshift errors, for both the testing and training samples. When the fit to the testing sample begins to worsen, the code stops changing the network.

\subsection{Empirical Method: Boosted Decision Trees}

We begin with a series of objects each described by a number of variables, we then define each object as either {\it signal} or {\it noise}. The objects within a predefined redshift bin will be labeled signal, and those outside this bin will be labeled noise. 

A decision tree works by iteratively dividing these objects into separate `nodes' based on a single variable at a time, where each node corresponds to a different region in parameter space. The point of division is chosen as the value which maximizes the separation between the signal and the noise. This division continues until a stopping criteria is achieved. The final nodes are labeled leafs, where each is classified as either signal or noise, finally the sum of these nodes is labeled a {\it tree}.

Decision trees can be sensitive to unphysical characteristics of the training data given they are unstable. This feature can be mitigated by the use of {\it boosting}, which works by re-weighting the objects that were previously mis-classified and then training a new tree \citep{Hastie08}. This allows one to generate multiple trees. A boosted decision tree (BDT) is formed using a weighted vote of all these trees, where the weight is computed from the mis-classification rate of each tree.   

To generate estimates of the photo-z PDF we use the boosted decision trees algorithm implemented in {\tt ANNz2} \citep{2015arXiv150700490S}. This code makes use of the machine learning methods implemented in the TMVA package4 \citep{Hoecker07}. Note other similar prediction trees algorithms have been introduced by \citet{Gerdes10} and \citet{CK13}. The advantage of BDTs over other machine learning algorithms is their simplicity and the speed at which they can be trained. Moreover, BDTs are one of the most effective methods to estimate photometric redshifts, as has been demonstrated in other comparisons \citep[e.g.][]{San14}.

For each {\rm class} $30$ different BDT are trained and then an optimal group is selected (the BDTs differ by the type of boosting, number of trees, etc.). An optimal group is found by ranking solutions with a separation parameter, this quantifies the level of distinction between the noise and signal. A PDF is then constructed from this ensemble of solutions, where each solutions is weighted and includes a contribution from the error estimated by each BDT.

To generate our results we include the following configuration settings: (1) The ratio of background (or noise) objects to signal objects is adjusted to be between 5 and 10, as a too large ratio will introduce a bias. (2) When defining the background sample for each redshift bin objects within the redshift interval where $\delta z  = z - z_{\rm bin}\sim 0.08$ are excluded.


\begin{figure*}
\centering
\includegraphics[clip,angle=270,width=\hsize]{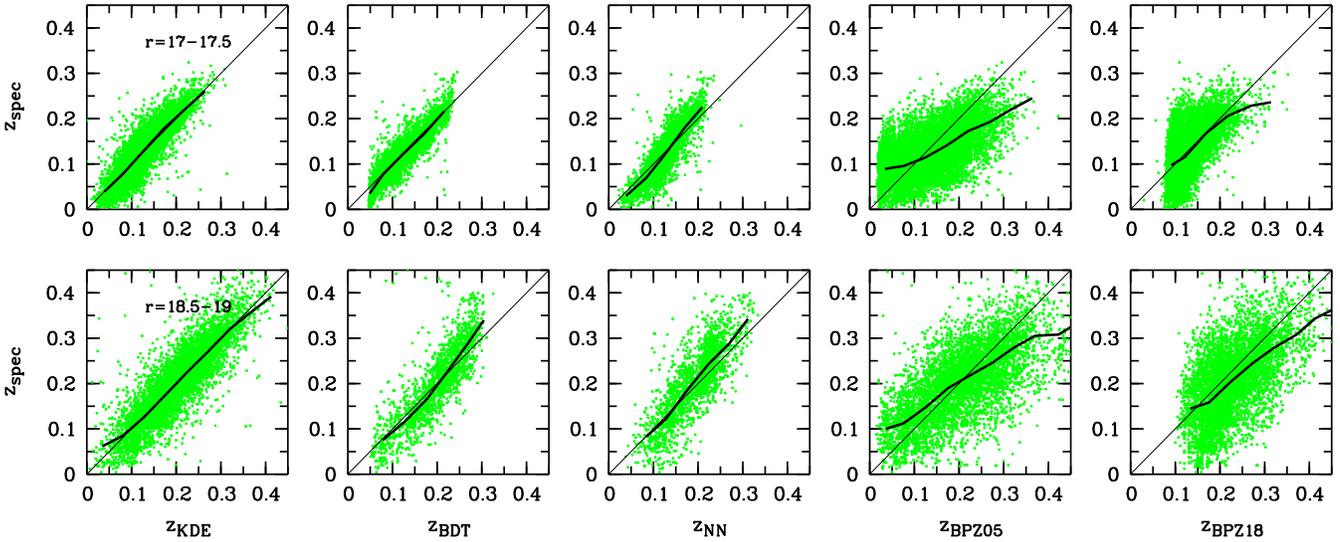}
\caption{Spectroscopic vs. photometric redshift in a bright (top) and faint (bottom) $r$-band magnitude bin. In this figure we draw attention to the very few outliers and redshift trends of the mean bias. {\it From left to right:} Kernel-Density Estimation (our code), Boosted Decision Trees (using ANNz2), Probabilistic Neural Net (using Skynet), Templates assuming zero-point uncertainty $0\fm05$ (using BPZ), and assuming $0\fm18$. Our template results (two columns on the right) show larger biases than the empirical methods (see discussion); they also could not use the WISE bands as these are not covered by the templates. 
\label{zz}}
\end{figure*}

\begin{figure*}
\centering
\includegraphics[clip,angle=270,width=0.73\hsize]{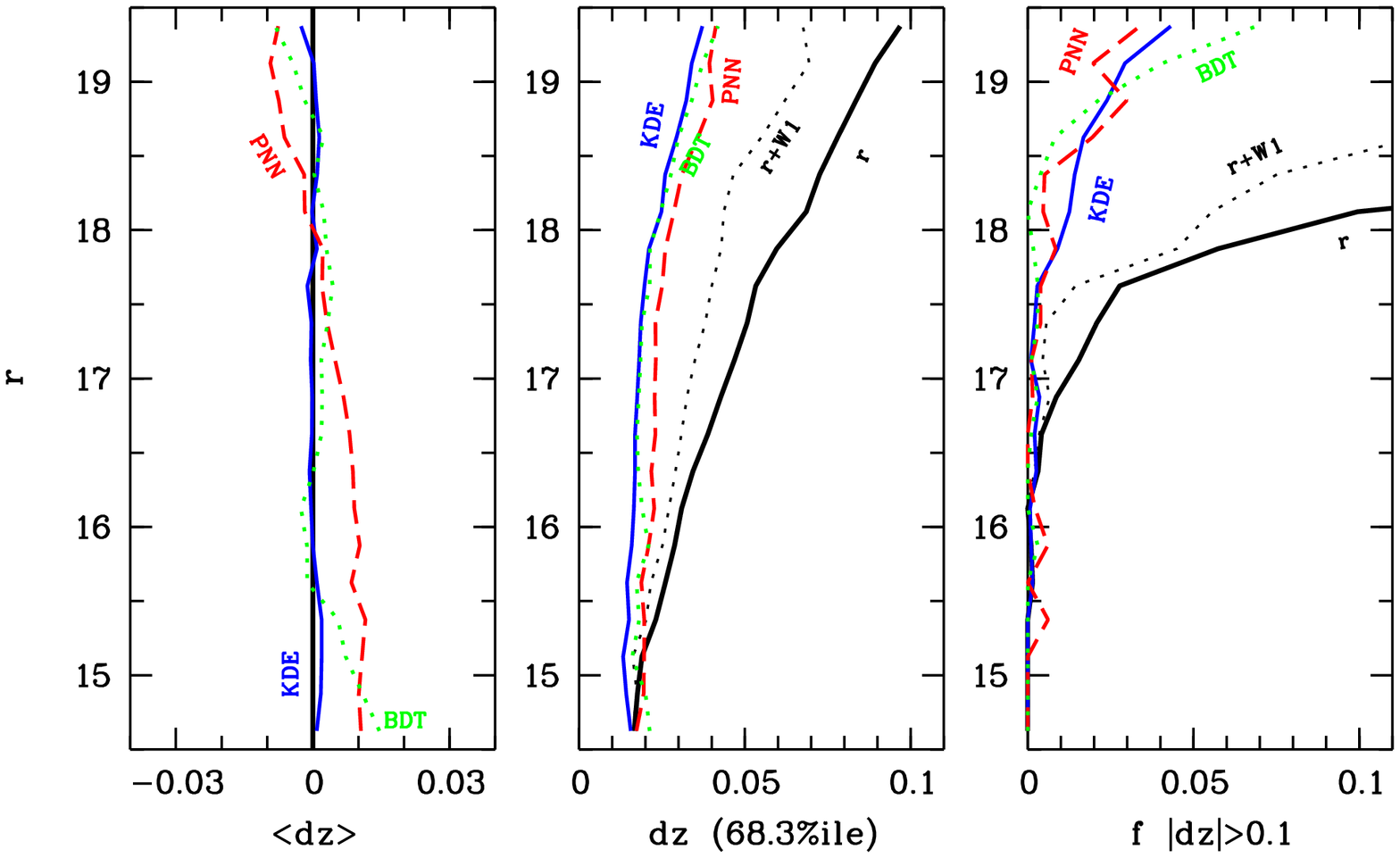}
\includegraphics[clip,angle=270,width=0.73\hsize]{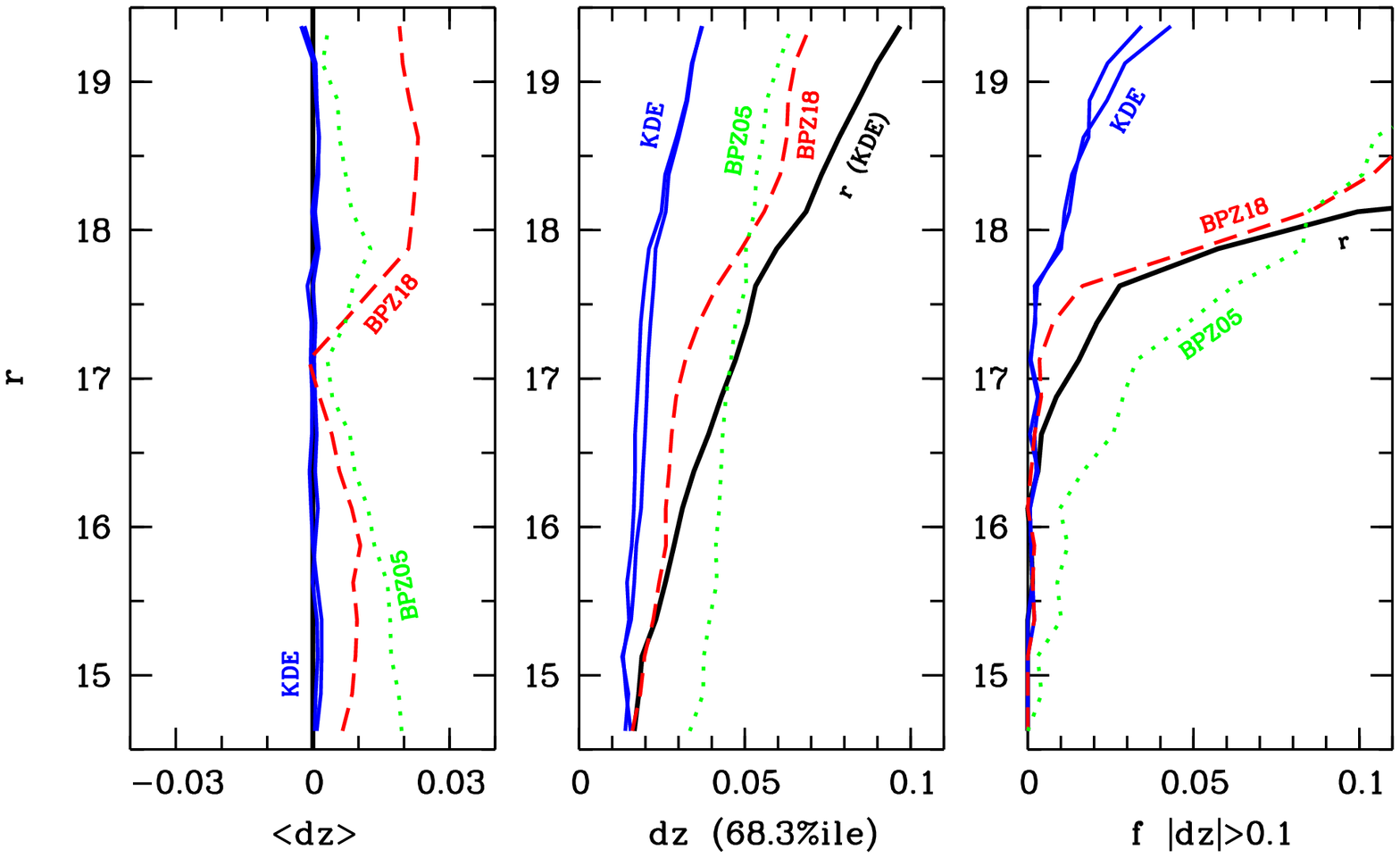}
\caption{Redshift statistics vs. $r$-band magnitude. {\it Left:} Mean redshift bias: intrinsic distribution (by definition zero, black solid line) vs. different methods. {\it Centre:} Half-width $\sigma_{683}$ of redshift deviation $\delta z/(1+z)$ containing 68.3\% of all objects. {\it Right:} Fraction of ''outliers'' with $|\delta z/(1+z)|>0.1$. {\it Top row:} Comparing the three different empirical methods and intrinsic redshift distribution (thick solid line labelled $r$). Note that the PNN method has excluded the $z>0.4$ galaxies, which would otherwise contribute to the outlier statistics. {\it Bottom:} Comparing the BPZ template method (using only $ugriz$) with the KDE method (two similar solid lines, one using all bands, one only $ugriz$) as well as the intrinsic redshift distribution (KDE using only $r$-band, thick line). 
\label{biasrms}}
\end{figure*}

\subsection{Bayesian template method BPZ}

As an example of a template-fitting method for photometric redshift derivation, we have chosen the Bayesian Photometric Redshift code \citep[BPZ,][]{B00,Coe06}, which is also the default method adopted by the KiDS survey \citep{deJong13}, although \citet{H16} relied on a spectroscopic recalibration of the obtained $p(z)$ estimates for the cosmic shear study in KiDS. BPZ applies Bayesian inference to estimate photometric redshifts by comparing broad-band photometry of a source with a set of redshifted template spectra. While the BPZ code is publicly available \citep{B11}, we used a slightly modified version which uses the numpy python package \citep{numpy} instead of the original numeric.

We used the re-calibrated template library of \citet{Cap04}, and a set of filters appropriate for the OmegaCAM instrument. Note that, unlike in the empirical approaches, we have not used the WISE photometric information as the templates do not cover the mid-infrared wavelengths.

In general, the BPZ code requires only a few parameters to be tuned, such as a bandpass used for the determining a redshift prior from the magnitude, the form of the prior itself, and the estimated uncertainty in the calibration zero point (on a band-to-band basis). As bandpass for the prior we used the $i$-band. The functional form of the default prior in BPZ was derived from the Hubble Deep Field North \citep[HDF-N, for details see][]{B00} and the Canada-France Redshift Survey \citep[CFRS,][]{Lilly95} and is optimised for high-redshift galaxies. Most of our galaxies are, however, located at relatively low redshifts, where the default prior performed badly. We thus adopted the prior by \citet{R14}, which takes into account the galaxy distribution at $i<20$ from VVDS \citep{VVDS} and is more appropriate for our data set.

We ran BPZ with several values for the zero-point uncertainty: initially, we used a fiducial estimate of $0\fm05$ (labelled BPZ05 in the following), but then varied the value as a free parameter and evaluated the results in terms of the mean bias and scatter of the photometric redshifts relative to the spectroscopic ones. We found that for our sample an uncertainty of $0.18$ gave on average optimal results, although it leads to other issues (see the discussion below).

\section{Results}

A first qualitative impression of the results is provided by Fig.~\ref{zz}, where spectroscopic redshifts are plotted vs. photometric ones for all four methods including the two versions of the BPZ template method. The two rows of the figure show a bright and a faint magnitude bin, with fainter galaxies obviously reaching higher redshifts. The closer the objects stay near the diagonal the better. The first obvious conclusion is that empirical methods hug the diagonal more closely than the template method, which is theoretically expected in the presence of rich and complete training sets. The difference between empirical and template methods is further explored in Fig.~\ref{biasrms} and Sect.~\ref{CompEmpTemp}, but first we discuss the results from the empirical methods themselves.

\subsection{Role of different passbands}

In this section, we first look at results obtained with a single method (KDE) but different sets of passbands. We assume that different empirical methods would find similar patterns for how results depend on data. In Tab.~\ref{rmsstats} we compare the redshift accuracy $\delta z/(1+z)$ in terms of the half-width of an interval containing 68.3\% of the sample, $\sigma_{683}$. The well-known factor $1+z$ accounts for the change in bandpass resolution with increasing redshift. We split the results into the bright sample from 2dFGRS ($r<17.7$) and the faint sample from 2dFLenS ($r=[17.7,19.5]$), as well as into red and blue galaxies using a rough bimodality cut defined by $g-r=0.5+2.8 z$ (based on the minimum between two modes in an observed-frame colour histogram after removing the mean slope with redshift). The empirical methods have two obvious trends in common:

\begin{enumerate}
 \item Bright objects have more accurate redshifts compared to faint objects.
 \item Red objects have more accurate redshifts compared to blue objects.
\end{enumerate}

Trend (i) makes clear that generic statements about the redshift accuracy of some photo-z method are meaningless, unless accompanied by a specification of the magnitude and the photometric signal-to-noise. At first sight, we may be tempted to consider (i) a consequence of photometric errors increasing for fainter objects. However, our relatively deep photometric data has nearly vanishing formal flux errors (see Tab.~\ref{tab:ATLAS_charac}), and hence our analysis takes place entirely within the saturation regime of photo-z quality \citep[for more details on error regimes see][]{WMR01}. Since magnitude has little effect on the photometric error ellipsoids in this work, any magnitude dependence of redshift errors must be driven by other factors; two possibilities are:

\begin{itemize}
 \item Calibration offsets between the model and data.
 \item The width of the intrinsic redshift distribution is a function of magnitude.
\end{itemize}

Given that we homogenised photometric zeropoints across our survey area, we don't expect model-data offsets in our empirical methods. However, we observe a strong trend in the redshift distribution with magnitude, both in the mean redshift and its scatter. We created a mapping of the form $z(r)$, based solely on the $r$-band magnitude while ignoring all SED information, and find that $\sigma_{683}$ increases from 0.017 at $r\sim 15$ to 0.1 at $r\sim 19.5$. The entire curve over five magnitudes of width is well fit by the relation $\log \sigma_{683} = -4.363+0.175 r$, i.e. per magnitude the true redshift scatter increases by a factor of $\sim1.5$. 

We simply implemented this process as a single-feature photo-z using the KDE formalism. For the subsamples, we find an average error of $\sim $0.04 in the bright sample and $\sim $0.08 in the faint sample, but of course these errors are simply a propagation of the width of the redshift distribution at fixed magnitude. Adding SED information to the process will shrink these errors, but in the faint sample we are starting from a wider base and don't expect to arrive at the same precision.

Trend (ii) has been observed for many years in photo-z's derived with only broad passbands using templates, where the stronger colour of red galaxies translates into a greater change of colour with redshift and hence a stronger signal\footnote{We note that medium-band filters tend to pick up emission lines as well \citep[e.g.][]{H94,WMR01,W04,W08,I09} and give blue objects an advantage they don't have in broadband data sets.}. However, the empirical methods show an additional influence from the intrinsic redshift distribution. The single-band photo-z's show again that at fixed magnitude red galaxies have a smaller redshift scatter than blue galaxies, even though both the red and the blue query sample have their redshifts estimated from the same overall map. In the fainter 2dFLenS sample, blue galaxies have an over 50\% wider intrinsic redshift scatter than blue galaxies, which is just the flip side of blue galaxies at fixed redshift showing a wider range of magnitudes, while red galaxies show a more peaked luminosity function \citep{Wolf03a,Bal04,Bell04}.

The most significant single-band addition we can make to the $r$-band is the W1 magnitude. At faint magnitudes where the intrinsic redshift distribution is wide, it helps reduce the redshift errors, e.g. from 0.08 to 0.055 in the overall 2dFLenS sample (using the same code). At $r\sim 15$, however, the intrinsic redshift distribution is narrowly concentrated to low redshifts of 0.025 to 0.06 (68th percentile), so there the W1 information is largely redundant with the $r$-band. Adding all bands to the process finally shrinks the redshift errors almost by half compared to $r$+W1 only. We note, that using the optical $ugriz$ bands without WISE photometry gives results that are 5 to 10\% worse than the full set including W1 and W2. This difference is small compared to those observed between some of the methods (see Tab.~\ref{rmsstats} and the following section).

At this point adding further broad passbands would largely add redundant information as there are no ambiguities to break, unless they probed additional sharp features outside the optical+WISE wavelength region. We thus conclude that additional improvements in redshift precision for galaxies at $r<20$ would mostly be enabled with filter sets that include narrower passbands to increase the sensitivity of colour measurements to smaller changes in redshift \citep{H94,WMR01,Spitler14,PAU,JPAS}, which routinely achieve very low bias and a redshift scatter of $\sim $0.007 for both galaxies and QSOs, even with template methods \citep{W04,W08,I09,B09}. 

\begin{table}
\caption{Redshift scatter $\sigma_{683}$, the 68th percentile of $\delta z/(1+z)$ by method and sample. Note that the BPZ code has used only the $ugriz$ bands. The mix of red:blue is 60:40 in the bright and 40:60 in the faint sample. \label{rmsstats} }
\begin{tabular}{lcccccc}
\hline  \hline  
			&  \multicolumn{3}{|c|}{$r<17.7$}	&  \multicolumn{3}{|c|}{$r=[17.7,19.5]$}	\\ 
Method		&  GRS		& red	&  blue	&  LenS 		&  red	&  blue	\\ 
\hline 
KDE $r$ only 	&  0.0401		&  0.037	&  0.045	&  0.0820		&  0.063	&  0.096	\\
KDE $r$+W1 	&  0.0338		&  0.033	&  0.034	&  0.0559		&  0.049	&  0.060	\\
KDE $ugriz$ 	&  0.0204		&  0.018	&  0.023	&  0.0310		&  0.025	&  0.035	\\
\hline
KDE all 		&  0.0179		&  0.016	&  0.021	&  0.0296		&  0.025	&  0.034	\\
BDT all 		&  0.0191		&  0.018	&  0.021	&  0.0315		&  0.027	&  0.035	\\
PNN all		&  0.0230		&  0.023	&  0.022	&  0.0354		&  0.032	&  0.037	\\
\hline
BPZ05 $ugriz$	 &  0.0454		&  0.038	&  0.057	&  0.0566		&  0.040	&  0.069	\\
BPZ18 $ugriz$	 &  0.0321		&  0.028	&  0.041	&  0.0613		&  0.047	&  0.070	\\
\hline
\end{tabular}
\end{table}

\subsection{Comparing different empirical methods}\label{CompEmpEmp}

There are two differences between the Kernel Density Estimation (KDE) method on the one hand, and the Boosted Decision Tree (BDT) and Probabilistic Neural Network (PNN) methods on the other: (i) The latter require a training procedure, and thus a split between training and validation sample, while KDE has no training process, and (ii) the trained methods didn't work at $z>0.4$, while for KDE we included in the model sample all galaxies up to $z=0.5$, which includes an additional $\sim $3\% of galaxies at faint magnitudes. In the query samples, we kept higher-redshift objects up to $z=0.6$ for KDE, above which all objects at this magnitude are QSOs that are identifiable in a photometric classification \citep[see e.g.][]{WMR01,W04,Saglia12,Kurcz16}.

In our comparison KDE is expected to be by far the slowest of all methods in terms of computer runtime, because it does not train a mapping, but calculates it on the fly during the estimation process. However, it leads in the precision of photo-z estimation in terms of both redshift bias and scatter (see Tab.~\ref{rmsstats} and Fig.~\ref{biasrms}), with BDT coming second in every single statistic, and PNN coming last in all of them. The difference is most apparent among red galaxies, where the redshift scatter in KDE is 20\% to 30\% lower than in PNN, while for blue galaxies the gain in KDE is only up to 10\%. BDT is consistently in the middle. We note that we find no significant redshift bias at all in the KDE method. The measured deviations are confined to approx. $\pm 0.001$ with the exception of the faintest quarter magnitude bin, and are consistent with the Poisson noise expected from the finite number of objects in a bin and their redshift scatter.

The lead of KDE in terms of minimised bias is theoretically expected. One variant of KDE has been shown by \citet{W09} to produce an exact frequency correspondence between query and model data, with zero bias and the only difference between true and estimated frequency being the propagation of Poisson noise arising from finite sample sizes. This variant of KDE is an exact implementation of Bayesian statistics with an empirical model sample, where the kernel function is chosen properly to take into account the feature errors in query and model objects leading to a ``zero-neighbourhood-smoothing KDE'' method. This approach takes into account that the original distribution of the model sample in feature space has already been smoothed by their photometric errors. In order to match it with the error-smoothed distribution of the query sample, the kernel smoothing should be restricted to the square \textit{difference} between query and model errors. However, the zero-smoothing approach requires that model errors are smaller than query errors, which is not the case in our study. Hence, we were only able to use a standard KDE method, which still comes close to zero bias. In Sect.~\ref{freqnz} we explore further to what extent the use of the standard KDE method implies differences between true redshift frequencies and those predicted here.

The BDT method performs nearly as well in terms of scatter, but less well in terms of bias. It demands more richly populated bins for training, and hence did not work for redshifts above $z=0.4$; and it shows a mild redshift bias at the low redshift end, and especially at higher redshifts towards $z=0.4$. Biases are expected when redshift is estimated via a classification approach working in fine redshift bins that become less fine where the training sample gets sparse. The redshift bins at lowest and highest redshift are the widest as the sample is sparsest there, and the need for training with such classes means that the full redshift resolution of the training set is not exploited. 

The PNN method performs slightly less accurately and is more biased at low and high redshifts. This is a modest disadvantage of NNs in the presence of rich training samples, where the smoothness of the map enforced by neural nets does not allow the full feature-space resolution of the training sample to express itself in the estimation process. However, the positive aspect of the smoothness criterion is that traditional regression NNs can outperform other methods when working with a sparse training sample, but this advantage is not expected to help with PNNs, where each redshift bin needs to be well-populated.

We note here a similarity with another method, not pursued in this work: Local Linear Regression (LLR) has been used by \citet{Csa03,Csa07} and \citet{Beck16} to derive photo-z's from SDSS photometry and empirical training sets. In this method a hyperplane is fitted to the redshifts of nearest-neighbour galaxies for each individual query galaxy, which implies and exploits a smooth (locally linear) colour-redshift relation to obtain more accurate redshift estimates even for locally sparse training samples. However, the construction of redshift distributions $p(z)$ does not benefit from this technique. 

\begin{figure}
\centering
\includegraphics[clip,angle=270,width=\hsize]{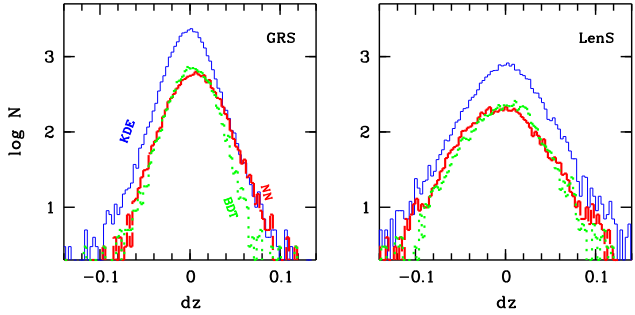}
\caption{Histogram of redshift deviations $\delta z/(1+z)$ for empirical methods. {\it Left:} Bright 2dFGRS sample. {\it Right:} Faint 2dFLenS sample. There is no sign of unusual outliers, which is expected for rich, complete, random training samples at $r<20$ and $z<0.6$, where no ambiguities appear in the colour-redshift map, although they contribute to the marginally richer-than-Gaussian tails of the distribution. 
\label{dzhisto}}
\end{figure}

The right-hand column of Fig.~\ref{biasrms} shows the rates of redshift ''outliers'', defined here as $|\delta z/(1+z)|>0.1$. These are very few objects, generally within less than 1\% at $r<18$, and increasing mildly at fainter levels. The methods appear comparable and their differences are largely due to statistical noise. It is important to clarify that these objects are not outliers in a classical sense, where true redshift ambiguities at one location in feature space leads to confusion. Instead, these objects are simply the wings of the error distribution and their fraction beyond a fixed threshold increases just as the width of the distribution increases. Fig.~\ref{dzhisto} shows the redshift error distributions of the three methods for the two samples, and proves that their distributions are nearly perfectly Gaussian apart from noise. We note that vertical offsets in Fig.~\ref{dzhisto} stem from the fact that KDE can consider every object a query object, while BDT and PNN need separate validation samples. This result is consistent with the general observation that at $r<20$ we observe a galaxy population confined to $z<0.6$ and without real ambiguities. 

However, this simplicity of the $r<20$ universe is in marked contrast to fainter data sets: at $r>20$ and $z>1$ significant ambiguities come into view, which then makes the consideration of redshift outliers worthwhile. In that different regime, we expect to see a difference in outlier handling by the KDE and BDT method: the boosting in BDT may act to suppress the propagation of unlikely signals into the trained map, as it is designed to suppress the propagation of rare false training signals. This would suppress the visibility of ``faint'' (high number-ratio) ambiguities in BDT, whereas KDE would take the model sample at face value. Thus BDT will produce cleaner results with messy training samples, while KDE will report fainter ambiguities than BDT provided the model samples are clean and trustworthy.

In summary, it appears that we have reached the intrinsic floor of redshift errors afforded by the data used in this study. The richness of the model sample has allowed KDE to outperform the neural nets and boosted decision trees and minimise the bias.

\begin{figure}
\centering
\includegraphics[clip,angle=270,width=\hsize]{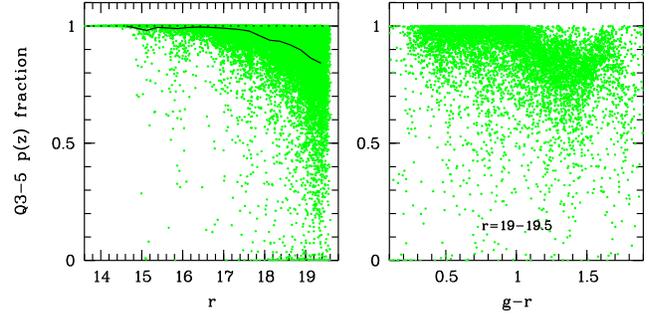}
\caption{Completeness of high-quality model as seen by the query sample of high-quality objects. {\it Left:} Per-object fraction of integrated frequency in high-Q sample (grey points) and mean trend (line). At $r>19$ there is a $\sim $15\% probability that an object's redshift is not drawn from the distribution of the high-Q sample defining the redshift map, but from the low-Q sample with unknown redshifts. {\it Right:} The high-Q fraction is lowest for faint red galaxies, whose spectra display only noisy absorption lines. 
\label{cplt}}
\end{figure}

\subsection{Spectroscopic completeness and residual estimation risk}

Spectroscopic incompleteness in an empirical model sample leads to a gap in our knowledge of the true $z(\vec{c})$ map, which implies that the estimated redshift probability distribution $p(z)$ is only part of the whole picture. While objects with unknown redshifts cannot be used to quantify redshift performance, they help at least to flag residual estimation risks due to this empirical gap. Is there a risk of getting photo-z's wrong when the spectroscopic sample is incomplete? The problem is that it is a-priori unclear, at what redshifts those objects reside, which were targeted by spectroscopy, but did not deliver a trustworthy redshift. The two possibilities here are: 
\begin{enumerate}
\item Those objects are a random subsample of objects with similar SEDs, so they reside at similar redshifts, and we have no redshifts simply because the signal in the spectrum was a little too weak
\item Those objects did not reveal their redshift precisely because they reside at redshifts different from the successful sample, which is why no strong features were seen that would have given away their redshift.
\end{enumerate}

These two alternatives have different implications for the photo-z misestimation risk: in case (1) the photo-z PDF will be correct after re-weighting, while in case (2) the shape of the photo-z PDF must be acknowledged to be wrong since the empirical method is blind to an important component of the redshift space. We don't know for sure, which case is realised here, but we have reasons to assume that at the moderate depth of $r\sim 19$ our incompleteness might be of the benign sort, as there is not much room for a significant fraction of objects to reside e.g. in the redshift desert at $z>1$. The situation is known to be different at fainter levels, where higher-z objects are selectively missing from spectroscopic samples \citep{DeepPC}.

In the KDE framework we can easily measure the probability $p_{\rm unknown}$ of any object to be attributed to the unknown redshift class. The mean $p_{\rm unknown}$ increases towards faint magnitudes in step with the spectroscopic incompleteness. The left panel in Fig.~\ref{cplt} shows that $p_{\rm unknown}$ increases significantly at $r>19$, up to $\sim $15\%. 

The right panel in Fig.~\ref{cplt} illustrates that redshift completeness is a function of galaxy colour. There appears to be almost a bimodality in the high-quality probability fraction of galaxies, such that red galaxies have systematically lower redshift completeness. This is expected since blue galaxies are star-forming and show a clear emission-line signature, which leads to high-confidence quality flags even for faint galaxies. The mean incompleteness for red galaxies with $g-r\approx 1.4$ and $r=[19,19.5]$ is $\sim $20\%, but some individual high-quality query objects fall into regions of the map, where the high-quality completeness goes to nearly zero.

\begin{table}
\caption{Outlier rates $f_{0.1}$ of KDE photo-z by spectroscopic quality flag Q. Rates are statistically consistent with the expected Q-dependent fraction of spectroscopic misidentifications.  \label{outliers} }
\begin{tabular}{lccrrrrr}
\hline  \hline  
Sample	&  mag range	& $\langle z\rangle$	&  Q4/5	&  Q3	&  Q2	&  Q1	\\ 
\hline 
2dFGRS	& $r<17.7$	& 0.111			&  0.1\%	&  1.7\%	&  13\%	&  -		\\
2dFLenS	& $r>17.7$	& 0.201			&  1.4\%	&  2.2\%	&  17\%	&  30\%	\\
 \hline
\end{tabular}
\end{table}

\subsection{Spectroscopic mistakes and outlier rates}

One of the reasons for the appearance of redshift outliers is errors in the spectroscopic identification, the rate of which will depend on the redshift quality flag assigned during the spectroscopic inspection. A quality flag of 4 or 5 indicates redshifts that seem absolutely certain to the human inspector. Accidental mistakes may still happen, but this quality class is expected to be notionally "99\% correct". With decreasing quality class we expect the fraction of incorrect redshifts to grow. $Q=1$ is assigned to all spectra where the human inspector cannot see any significant feature, and the redshift is taken from a cross-correlation fit. The bright 2dFGRS sample contains no objects with $Q=1$, but 3\% of the fainter 2dFLenS sample fall into this category.

We define outliers using $|\delta z/(1+z)|>0.1$ and measure their fraction among KDE photo-z's after splitting the sample by quality flag. Table~\ref{outliers} shows that outlier rates are lower for 2dFGRS than for 2dFLenS. This is a consequence of their intrinsic difference with mean redshifts of $\langle z_{\rm GRS} \rangle = 0.11$ vs. $\langle z_{\rm LenS} \rangle = 0.20$. Outliers as defined here, with a redshift error greater than 0.1, are almost ruled out by definition in a bright sample at consistently low redshift. 

Among Q=4 and 5 objects we find outlier rates on the order of $\sim $1\%, which is fully consistent with the notional target reliability of 99\% for this quality class. We checked the hypothesis that those outliers might be due to wrong spectroscopic redshifts rather than wrong photo-z's. Revisiting their spectra, we found that less than 20\% of these conflicting cases had uncertain or wrong redshifts assigned. We have not yet investigated in detail what explains such outliers that are not expected in the absence of ambiguities in the colour-redshift map. However, we have evidence that source blending and strong lensing play a role, where two galaxies at different redshifts appear on nearly the same line-of-sight. This affects the measured photometry of the combined object and makes the redshift itself ambiguous.

We also find that the outlier rates increase with decreasing spectroscopic quality, as expected. They reach $\sim $2\% for Q=3, $\sim $15\% for Q=2 and $\sim $30\% for Q=1. It appears credible that the increases above the rate for the highest-quality bin are mostly due to spectroscopic data quality issues. Bad seeing, high background and faint objects all lead to lower signal-to-noise, and scattered light from nearby objects confuses the situation. Spectroscopic quality decreases for fainter objects and generally for worse observing conditions. However, there do exist galaxy spectra, which are intrinsically challenging due to weak lines, e.g. when low metallicity is combined with low star-formation rate. We repeat, however, that in our magnitude range we do not expect any galaxies from the redshift desert of optical spectrographs at $z\ga 1$.

The spectroscopic redshift of objects with Q=1 is almost uniquely determined by cross-correlation of the spectrum with templates, and specifically marks cases with a very weak correlation coefficient. A $\sim $30\% outlier rate above a threshold of $|\delta z/(1+z)|>0.1$ is consistent with either the photo-z or the spec-z being drawn at random from the full sample, while the other one is assumed to be correct. So, this could mean that the Q=1 cross-correlation redshifts are random, while the photo-z's are actually correct, or the photo-z's are random while the low cross-correlation redshifts are correct, or any mix of the two. Only for Q=2 and better quality levels, a human assessor has actually labelled our confidence in the spectroscopic redshift as "possible" (Q=2), "probable" (Q=3), and "practically certain" (Q=4/5).

\begin{figure}
\centering
\includegraphics[clip,angle=270,width=\hsize]{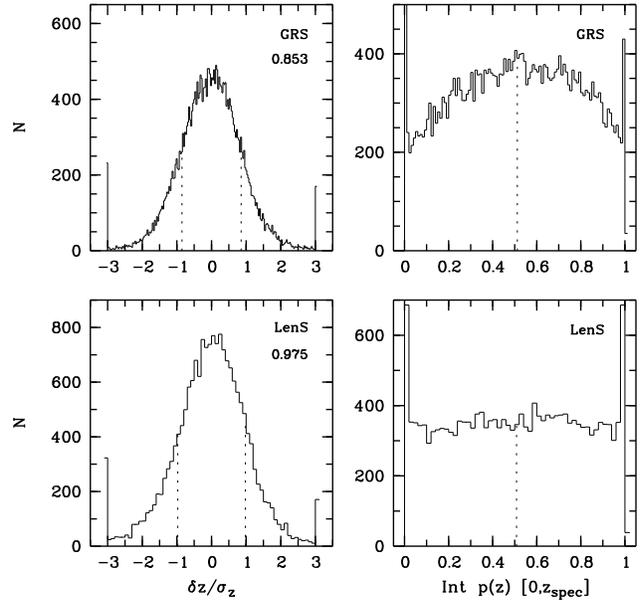}
\caption{Validity of per-object estimated redshift probability distributions. {\it Left:} Ratio of estimated rms and true redshift rms. {\it Right:} Distribution of true redshifts within estimated redshift distribution. The kernel smoothing in standard KDE makes estimated redshift distributions wider than true redshift distributions, so that true redshifts are not randomly drawn from the estimated distribution. This effect is noticeable for the brighter objects in 2dFGRS, but negligible for the fainter 2dFLenS objects. The excess in the histogram bins at both ends of each distribution are redshift outliers. 
\label{relerrors}}
\end{figure}

\subsection{Frequentist interpretation of $p(z)$}\label{freqnz}

It is often asked whether photometric redshift probability distributions represent actual redshift frequencies for objects of the colour they were derived for. If this was the case, we could choose any subsample and add their $p(z)$ to obtain a realistic $n(z)$ estimate. \citet{W09} have shown that in the limit of zero-smoothing KDE such estimates deviate from the true $n(z)$ only by Poisson noise. However, zero-smoothing KDE can only be implemented if the model photometry is less noisy than the query photometry. In this work model and query objects have statistically the same errors, which in zero-smoothing KDE implies a kernel that is a $\delta$-function. Choosing a kernel of finite width to avoid a breakdown of the method resulting from the discrete nature of the model applies additional smoothing to the $z(\vec{c})$. This kernel smoothing introduces neighbourhood information into the model where it is not expected in the data. The effect is expected to be strongest for bright objects, while the distribution of faint objects in feature space is already smoothed by larger photometric errors so that additional smoothing has less effect. We use two diagnostics to test whether $p(z)$ statistically represents $n(z)$. 

First we determine the rms redshift error $\sigma_z$ expected from the $p(z)$ and compare it to the true redshift error $\delta z$. For an accurate $p(z)$ we expect the ratio of $\delta z/\sigma_z$ to have zero mean and rms $=1$. In the left-hand panels of Fig.~\ref{relerrors} we show the distribution for the bright sample with $r<17.7$ from 2dFGRS (top row), and that for the faint sample at $17.7<r<19.5$ from 2dFLenS (bottom row). We first note the presence of $>3\sigma$-outliers, where the true redshift is much further from the estimate than the width of the $p(z)$ distribution suggests. Outliers were discussed in the previous section. We then see that the shape of the distribution is approximately Gaussian with an rms of 0.853 for 2dFGRS and 0.975 for 2dFLenS. So, the width of the $p(z)$ distribution seems on average appropriate for the fainter 2dFLenS sample, while the kernel smoothing overestimates the expected errors for its redshift estimates of the bright 2dFGRS sample, just as expected theoretically.

Now we probe the shape of the $p(z)$ distribution: we check whether the spectroscopic redshifts $z_i$ are randomly drawn from $p_i (z)$ on a per-object basis. For this purpose, we measure the fraction of the normalised estimated probability below the true redshift:

\begin{equation}
	 {\int_0^{z_{\rm spec}} p(z) dz} \Big/ { \int_0^{z_{\rm max}} p(z) dz} 	~.
\end{equation}

If the $p(z)$ is correct, we expect the distribution of the fractional probability to be precisely flat apart from noise. The right column in Fig.~\ref{relerrors} shows that this is indeed the case for 2dFLenS (bottom right panel): the distribution is flat and the median value of the sample is $0.508$, very close to the expected value of $0.5$. If we clip the outlier bins (see below), we find mean numbers per bin of $\sim $348 and an rms of $\sim $20, very close to Poisson noise. Any smaller sub-sample selected from the whole 2dFLenS sample would be even more dominated by Poisson noise, so that possible systematics would be even harder to detect. We thus take it as established that the KDE $p(z)$ estimates in any subsample at $r\ga 18$ from our dataset are consistent with true redshift frequencies plus Poisson noise.

The brighter 2dFGRS sample shows a similar median value, but the sources are found to be concentrated more toward the centre of their estimated distribution than suggested by the distribution itself. This corroborates the previous finding and theoretical expectation that the kernel smoothing has added excess probability to the wings of the expected distribution for bright objects. 

Finally, the bins at the ends of the distributions mark objects whose redshift is not within the central 96th percentile of their estimated redshift distribution, and the excess in those bins amounts to $\sim $700 objects or a $\sim $3.5\% fraction among the total 2dFLenS sample and $\sim $500 objects or $\sim $1.5\% of the 2dFGRS sample. These objects are clear outliers from their estimated $p(z)$, and may be due to source blending.

\subsection{Comparing template methods with empirical methods}\label{CompEmpTemp}

We first note that template methods have more room for mismatches between model and data than empirical methods, which lead to redshift biases when left uncorrected. We note the following issues: 

\begin{enumerate}
 \item Templates may not reliably cover all the features probed by the observations because their physics may not be understood and calibrated well enough to make a parametric model that is bias-free. E.g, we did not use the WISE W1 and W2 bands in the template comparison as the templates are unreliable at this wavelength. We may expect a little less redshift discrimination in the BPZ results, but note that excluding WISE made less than a 10\% difference to the KDE results.
 \item Redshift priors in template methods depend on the chosen template grid and will be sub-optimal given that template grids are a simplified view of the true SED space occupied by real galaxies. We found that the default prior of BPZ was inappropriate for $r<20$ objects, as it had been tuned for fainter galaxies in deep surveys, so we chose a different prior from the literature, which may still not be a perfect match.
 \item Calibration mismatches between templates and photometric data have been reported fairly often, and can be addressed with photometric zero-point adjustments in the observed frame as well as with template repair methods in the restframe \citep[e.g.][]{Csa00,Bud00,Bud01,Csa03}. 
\end{enumerate}

Fixing any of these three issues is beyond the scope of this paper, and thus our results for the template method remain probably sub-optimal. As we see in Fig.~\ref{zz} and Fig.~\ref{biasrms}, the most obvious difference between our photo-z results from template methods and empirical methods are redshift biases that are a function of redshift.

In contrast to our application of the template method, an empirical method can make use of all features observed in the data set. When it is based on a random sample drawn from a large volume in a homogeneous Universe observed with homogeneous photometric zero-points, it will also naturally take care of the calibration and the prior of the model. Everyone with a rich complete training set has the option to use an empirical method, unless extrapolation in magnitude, redshift or object type are required, which would then involve uncertain parametric assumptions. 

\begin{figure}
\centering
\includegraphics[clip,angle=270,width=\hsize]{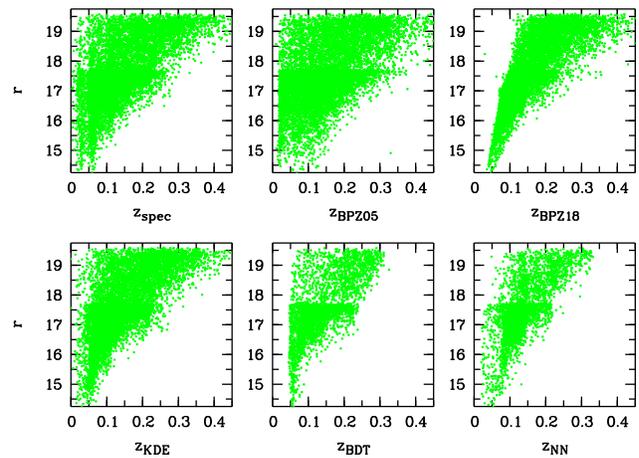}
\caption{Redshifts vs. magnitude. The focus of this figure is on regions that are not populated. {\it Top left:} Spectroscopic redshift. {\it Other panels:} Photo-z by method. In BPZ18 photo-z's avoid low redshifts for fainter magnitudes as the wide minimum error and the steep redshift prior drive the estimates higher. BDT and PNN are affected by the sparsity-induced upper redshift limit of $z=0.4$. 
\label{zR}}
\end{figure}

A further source of redshift biases is the profound semantic difference between a likelihood and a probability. The estimated zero-point error plays a special role in template methods: it is usually added quadratically to an object's flux error to represent a more realistic uncertainty in an SED. This widens the Bayesian likelihood contours and protects the user against trusting optimistic error estimates. However, it is clear that the $p(z)$ functions thus obtained are likelihood estimates and not probability estimates, representing degrees of belief, and not frequencies in a random experiment. We repeat that in our case the formal flux errors are nearly vanishing and the assumed error is basically the zero-point error.

We first assumed fiducially a zero-point uncertainty of $0\fm05$, but noticed that at $r<17$ a single-band redshift estimated empirically from only the $r$-band magnitude is already more accurate than the five-band template method BPZ05 (see Fig.~\ref{biasrms}). We re-ran the code with other values and found that bias and scatter were on average reduced by increasing the zero-point uncertainty to $0\fm18$. 

However, this change also introduced specific features into the results as the inflated flux confidence intervals then corresponded to a broader range of likely redshifts. Near the edges of the redshift distribution, where the redshift priors are steep, intermediate redshifts are thus rendered more likely than the correct very low or very high redshifts. This effect is shown in Fig.~\ref{zR}, where we show the redshift estimates of the entire sample vs. $r$-band magnitude. While the estimates of $z_{\rm BPZ05}$ (with $0\fm05$ zero-point uncertainty) cover the entire space of realised redshift-magnitude realisations, those with the inflated zero-point uncertainty, $z_{\rm BPZ18}$, are confined to a much tighter redshift interval at fixed magnitude and are systematically devoid of low-redshift estimates with increasing magnitude, such that e.g. no single photo-z for $z<0.1$ occurs at $r\sim 19$. But we note, that photo-z-based science applications often require that the true redshift distribution in a photo-z bin is correct, and not the converse, which would not be helpful given that in a blind photo-z situation, there is obviously no possibility to sort objects into true redshift bins. This is indeed the case for $z_{\rm BPZ18}$ at the low-redshift end, while it is not the case for $z_{\rm BPZ05}$ at either end.

\section{Conclusion}

In this paper, we present a new training set for photometric redshifts of galaxies, which is complete and randomly selected to $r\la19.5$, and thus two magnitudes deeper than the SDSS Main Spectroscopic Galaxy sample. It is similarly deep as the GAMA redshift sample, but randomly sub-selected from a larger survey footprint to suppress large-scale structure. One new aspect of this training sample is that we counteracted the steep number counts of galaxies by selecting brighter galaxies from a larger effective area within the same survey footprint. This reduces redundancy of information at fainter magnitudes and simultaneously reduces sparsity of representation and model discretisation noise at brighter magnitudes. Together with the 2dFGRS sample at brighter magnitudes, it includes over 50,000 redshifts selected from over 700 deg$^2$ area in the Southern sky.

Based on this training set and $ugriz$ and WISE photometry, we investigate the performance of photometric redshifts for galaxies using empirical and template approaches. All our methods are set up so that they produce redshift probability distributions $p(z)$ and not simply redshift estimates in the first instance. We investigate both the comparison of best estimates derived from the $p(z)$ as well as the precision of the $p(z)$ in actually representing expected frequencies.

We find that the template method suffers from redshift biases due to several issues that are not easy to fix. The empirical methods have lower biases, especially the KDE method, which by design has no measurable bias larger than Poisson noise. None of the methods have significant outliers, since there are simply no serious colour-redshift ambiguities at $r\la 20$ and $z<0.6$, as opposed to deeper and higher-redshift surveys. 

Two of our empirical methods, using boosted decision trees and neural nets, are trained methods and demand sufficiently dense training samples. Thus they tend to fail at the edges of the redshift distribution, where the number counts are steep and the sample becomes sparse. In contrast, the KDE method is not a trained method, hence our spectroscopic sample acts directly as a model set instead of a training set, as it is simply a discrete random representation of the galaxy redshift density field in observed feature space. 

The KDE method also achieves the best photometric redshift scatter, with values for the 68th percentile of $\sigma_{683}=$(0.014, 0.017, 0.028) for samples selected to have $r<(15.5, 17.5, 19.5)$, respectively. We note that additional improvements in redshift precision for galaxies at $r<20$ would mostly be enabled with filter sets that include narrower passbands to increase the sensitivity of colour measurements to smaller changes in redshift, but are not expected from adding further, largely redundant, broad-band information.

We reiterate that photometric redshift precision is necessarily a strong function of magnitude, since intrinsic redshift variance itself is a strong function of magnitude, increasing from 0.017 at $r\sim 15$ to 0.1 at $r\sim 19.5$. Often also the signal-to-noise of the photometric feature measurement is a strong function of magnitude, although this is not the case in our data set, where photometry is uniformly good across the entire range considered here. As a result, many statements in the literature about photometric redshift scatter contain no information unless accompanied by clear magnitude and signal-to-noise references. 

We note, however, that the realisation of the KDE method presented here is still not a textbook realisation of Bayesian statistics, as such an implementation requires photometric errors to be smaller on the model sample than on the query sample. Only then can we implement KDE in such a way that $p(z)$ represents the expected redshift frequency precisely. However, at $r>18$ we don't see a measurable deviation from the expected frequencies plus Poisson noise for any subsample of objects.

In summary it appears that photometric redshifts for galaxies to $r\la 20$ are a solved problem, and the next frontier is improving model samples at $r>20$ by making them simultaneously more complete, random and less affected by large-scale structure. A helpful project to this end, scheduled for observation in the 2020s, is the Wide Area VISTA Extragalactic Survey \citep[WAVES,][]{WAVES} planned at the forthcoming 4MOST instrument at the ESO-VISTA telescope, which may end up creating a large and complete random redshift sample to $r=22$.
 
The original motivation for obtaining this training set was to use it for deriving KDE-based photometric redshifts in the SkyMapper Southern Survey, which will see a first release of its Main Survey data in 2017. At $z<0.2$, we expect the redshift scatter to be slightly better for the SkyMapper filter set than for $ugriz$, because of the narrower $uv$ (ultraviolet and violet) filter pair in SkyMapper. For SkyMapper, we also plan a KDE classification into object types, which we will explore with further training sets that we have obtained in the realm of stars and quasars.

\section*{acknowledgements}
CW was supported by Australian Research Council Laureate Grant FL0992131.
Parts of this research were conducted by the Australian Research Council Centre of Excellence for All-sky Astrophysics (CAASTRO), through project number CE110001020.
MB is supported by the Netherlands Organization for Scientific Research, NWO, through grant number 614.001.451, through FP7 grant number 279396 from the European Research Council, and by the Polish National Science Center under contract \#UMO-2012/07/D/ST9/02785.
CB acknowledges the support of the Australian Research Council through the award of a Future Fellowship. 
CH acknowledges support from the European Research Council under grant number 647112. HH was supported by the Deutsche Forschungsgemeinschaft under Emmy Noether grant Hi 1495/2-1. TE and DK are supported by the Deutsche Forschungsgemeinschaft in the framework of the TR33 `The Dark Universe'. 
Based in part on data acquired at the Australian Astronomical Observatory, through program A/2014B/08.
KK acknowledges support by the Alexander von Humboldt Foundation.
DP is supported by the Australian Research Council Future Fellowship Grant FT130101086.
This publication used data obtained by the ESO VST-ATLAS survey under program ID 177.A-3011. It also makes use of data products from the Wide-field Infrared Survey Explorer, which is a joint project of the University of California, Los Angeles, and the Jet Propulsion Laboratory/California Institute of Technology, funded by the National Aeronautics and Space Administration. 

\end{document}